\documentclass[12pt,a4paper]{article} 
\usepackage[dvipsnames,usenames]{color}
\usepackage{graphicx}
\usepackage[normalem]{ulem}
\usepackage{lscape}
\usepackage{multirow}
\usepackage{ftnxtra}

\begin{document}

\title{OV or TOV ?}

\author{\.{I}brahim 
Semiz\thanks{mail: ibrahim.semiz@boun.edu.tr} \\ \\
Bo\u{g}azi\c{c}i University, Department of Physics\\
Bebek, \.{I}stanbul, TURKEY}
    
\date{ }

\maketitle

\begin{abstract}
The well-known equation for hydrostatic equilibrium in a static spherically symmetric spacetime supported by an isotropic perfect fluid is referred to as the Oppenheimer-Volkoff (OV) equation or the Tolman-Oppenheimer-Volkoff  (TOV) equation in various General Relativity textbooks or research papers. We scrutinize the relevant original publications to argue that the former is the more appropriate terminology.
\end{abstract}


\section{Introduction}
\label{intro}

The concept of a perfect fluid, one with no viscosity or heat conduction, is an idealization that serves as a simplifying assumption in many problems, both in pre-relativistic and relativistic phyics. Despite being an approximation, it can lead to quite realistic results in appropriate contexts. At the down-to-Earth scale, the ideal gases of thermodynamics qualify as perfect fluids, and at the very large scale, the universe is taken to be filled with a cosmic perfect fluid whose ``atoms'' are the galaxies. At the intermediate scale, the plasma that makes up stars also behaves as a perfect fluid to a good approximation, and the same is assumed for stellar bodies other than regular stars, i.e. white dwarfs and neutron stars.

While small celestial objects (asteroids or small satellites) can have irregular shapes, large ones are round. The reason is quite intuitive: Deviation from spherical shape will include regions higher than the average radius, that is, mountains; but a mountain will tend to spread due to its own weight. On small solid (rocky and/or icy) objects, gravity is weak, so the strength of the materials can withstand it, but on objects larger than a few thousand km in diameter, gravity working over millions of years will smooth any mountain or depression.

A celestial object made of a perfect fluid, a ``star'', by definition will not be able to support shear stresses, therefore mountains; hence is expected to be spherical if static, a very intuitive argument for sphericity. Yet it has proven surprisingly difficult to rigorously prove spherical symmetry of static perfect fluid stars. For Newtonian gravity, the proof was given  in 1919  \cite{Carleman,Lichtenstein}; for General Relativity (GR), a complete proof still does not exist. The conjecture was first explicitly stated in 1955 \cite{Lichnerowicz}, proven for a certain class of equations-of-state (EoS -- pressure-density relationships) in 2007 \cite{Masood}, and for a wider class in 2011 \cite{Pfister}.

When one makes the assumption of spherical symmetry in GR, the ansatz for the spacetime metric simplifies greatly, hence, so do the Einstein's equations. The problem of finding exact solutions for the spacetime metric of such a perfect-fluid body was first addressed in 1939, in back-to-back papers published in {\it Physical Review} by Tolman \cite{Tolman} and Oppenheimer \& Volkoff \cite{OV}, hence the relevant equation, to be rederived in  section \ref{sect:OV}, is called the Oppenheimer-Volkoff (OV) or Tolman-Oppenheimer-Volkoff (TOV) equation. The question of the appropriate choice is the subject of the present paper.

Of the two dominant textbooks of General Relativity published in early seventies, ``MTW'' \cite{mtw} calls the equation [eq.(23.22) of that book] OV, 
the other, Weinberg \cite{WeinbergGC}, diplays the equation [eq.(11.1.13) of that book] without giving it a name, or providing a reference [It {\it does} use the term ``Oppenheimer-Volkoff limit'' in a different, but related context, though]. The dominant GR textbook of the next decade, Wald \cite{WaldBk}, calls it [eq.(6.2.19) of that book] the TOV equation, as does the quite recent tome of Zee \cite{ZeeNutShellGR} [eq.(13) in Sect.VII.4 of that book]. 

In the research literature, the first use  of the equation we could find after 1939 is in 1959 \cite{Cameron}, without giving the equation a name, but crediting Oppenheimer \& Volkoff \cite{OV}; the use in \cite{arm} and \cite{BellThesis} in 1961 [eq.(2.3) and Ch.5, eq.(5) of the respective works] is similar. The work \cite{MisnerSharp} in 1964 is the first to use the phrase ``Oppenheimer-Volkoff equations of hydrostatic equilibrium'' but does not display the equation(s). The review article \cite{fowler} in the same year displays the equation, without assigning a name to it, and only very indirectly crediting Oppenheimer \& Volkoff \cite{OV} (but not Tolman); and a report \cite{UCalReport} credits them explicitly. Starting in 1965, such references increase in frequency; and in 1968, the first papers appear that both display the equation, and refer to it by name; except that in two papers, Tolman's name is also associated with the equation: The work \cite{arm2} calls a set of displayed equations the ``Oppenheimer-Volkoff equation'', whereas \cite{HartleThorne} calls it the ``Tolman-Oppenheimer-Volkoff equation of hydrostatic equilibrium'' and \cite{Bludman}, the ``Tolman-Oppenheimer-Volkoff condition''. In the same year, \cite{Wheeler68} uses the phrase ``Tolman-Oppenheimer-Volkoff general relativistic equation of hydrostatic equilibrium'' without displaying the equation.

In 1969, we could find no use of the phrase ``OV equation'' or equivalent, and two uses of ``TOV equation'' or equivalent. The corresponding numbers are one and one for 1970, two and two for 1971, none and four for 1972, and none and four for 1973. We end the year-by-year breakdown with 1973, the year of publication of the influential  book ``MTW'', and give a decade-by-decade breakdown in Table \ref{table:OVvsTOV}\footnote{The searches were performed using Google Scholar (GS). For the first row, the phrase ``Oppenheimer-Volkoff'' was searched for, the expressions ``Oppenheimer-Volkoff limit'', and of course ``Tolman-Oppenheimer-Volkoff'' were explicitly vetoed, and the appropriate uses of the expression were counted by visually inspecting the GS blurbs; the numbers for the last two decades were estimated as roughly 75\% of the total count, based on experience with previous decades. For the second row, the exact expression was searched, which left out some misspellings (or misidentifications of GS's OCR software) of Tolman, and a few instances of ``Oppenheimer-Volkoff-Tolman'' that we had encountered during the scrutiny for the first row; and added very few extras like ``Tolman-Oppenheimer-Volkoff solution''; therefore should be approximately correct.} .
We also note that one of the authors of that appropriately gravitating tome used ``OV'' in a previous publication \cite{MisnerSharp} and two of them ``TOV'' in a publication and a talk \cite{HartleThorne,WheelerTalk}, yet in the book they used ``OV''. On the other hand, the table tells us that the use TOV has always been more popular, but started to really dominate in the last two decades.

\begin{table}[h!]
\begin{center}
\begin{tabular}{ | c  || c  | c | c | c | c | c |} \hline
\multirow{2}{*}{{\bf Name}} & \multicolumn{6}{c|}{{\bf Decade}}  \\  \cline{2-7}

& 60's & 70's & 80's  & 90's & 00's & 10's\\  \hline \hline
OV  & 2 & 11 & 51 & 102 & $\sim$ 200 & $\sim$ 140 \\  \hline
TOV & 6 & 48 & 52 & 181 & 830 & 1420   \\  \hline
\end{tabular}
\end{center}
\caption{Approximate numbers of occurence of the expressions ``Oppenheimer-Volkoff equation'' (and equivalents) and ``Tolman-Oppenheimer-Volkoff'' in the research literature in each decade since the 1960's.} 
\label{table:OVvsTOV}
\end{table}

To discuss the question of more appropriate usage, in the next section we describe the setting of the problem. In Section \ref{sect:Tolman}, we review and discuss the first of the relevant papers \cite{Tolman}, by Tolman, and in Section \ref{sect:OV}, we review and discuss the second one \cite{OV}, by Oppenheimer \& Volkoff. In the final section, we evaluate and conclude.


\section{The Setting}
\label{sect:setting}

  The question at hand is finding valid solutions for the contents and structure of a static spherically symmetric  spacetime filled with an isotropic perfect fluid.  Of course, the spacetime structure is assumed to be described by a metric, which obeys Einstein's Field Equations (EFE)
\begin{equation}
G_{\mu\nu} = \kappa T_{\mu\nu}
\label{EFE}
\end{equation}
where $G_{\mu\nu}$ is the Einstein tensor, $T_{\mu\nu}$ the stress-energy-momentum (SEM) tensor, and $\kappa$ the coupling constant, including Newton's constant $G$. The relation between $G_{\mu\nu}$ and the metric, $g_{\mu\nu}$,
 is given in any GR textbook, e.g. \cite{mtw}, and we use that book's sign conventions for the definitions of the intermediate mathematical objects, that is, the Riemann and Ricci tensors, and the Christoffel symbols.
 
The most general form for the line element of a static spherically symmetric spacetime is
\begin{equation}
ds^{2} = -B(r) dt^{2} + A(r) dr^{2} + r^{2} d\Omega^{2}   \label{ansatz}
\end{equation}
where $d\Omega^{2} = d\theta^{2} + \sin^{2}\theta \, d\phi^{2}
$ is the metric of a two-sphere; and a perfect fluid is characterized by a  stress-energy-momentum tensor of the form
\begin{equation}
T_{\mu\nu} = (\rho + p) u_{\mu} u_{\nu} + p g_{\mu\nu}  \label{pfemt}
\end{equation}
where $\rho$ and $p$ are the energy density and pressure, respectively, as measured by an observer moving with the fluid, and $u_{\mu}$ is its four-velocity; in units such that the speed of light $c$=1. Since the fluid is  at rest, we have
\begin{equation}
u^{\mu} = u^{0} \delta_{0}^{\mu}	\label{staticu}
\end{equation}
and of course, the four-velocity is normalized such that  
\begin{equation}
u_{\mu}u^{\mu} = -1 	\label{unorm}
\end{equation}

After these preliminaries, the Einstein Equations can be written down. The nontrivial components (00, 11 and 22) are
\begin{eqnarray}
\frac{B}{r^{2}} \left( 1- \frac{1}{A} + \frac{rA'}{A^{2}} \right)  & = & \kappa B \rho(r) \label{EFE00} \\ 
\frac{1}{r^{2}} \left(1- A + \frac{rB'}{B}\right) & = & \kappa A p(r) \label{EFE11} \\ 
\frac{r}{2A} \left[ -\frac{A'}{A} + \frac{B'}{B} - \frac{rA'B'}{2 AB} - \frac{rB'^{2}}{2 B^{2}} + \frac{rB''}{B} \right] & = & \kappa r^{2} p(r) \label{EFE22}
\end{eqnarray}
where $A(r)$ and $B(r)$ are written as $A$ and $B$ for brevity, and prime denotes $r$-derivative (The 33 component is simply the 22 component multiplied on both sides by $\sin^{2}\theta$).

Alternatively, these equations can be combined to give
\begin{equation}
\frac{p'}{\rho+p} + 2 \frac{B'}{B} = 0 	\label{B-p-rho}
\end{equation}
which can be used instead of the complicated 22 component, eq.(\ref{EFE22}). This last equation (where we stopped explicitly showing the $r$-dependences of $p$ and $\rho$, as well) could also have been derived from the local energy-momentum conservation equation, $T^{\mu\nu}_{\;\;\;\; ;\nu} = 0$, which in turn follows from the mathematical fact $G^{\mu\nu}_{\;\;\;\; ;\nu} = 0$ (the ``contracted Bianchi identity'') and the Einstein Equations (\ref{EFE}). This is the well-known statement that in GR, local energy-momentum conservation is built in.

Whichever set one chooses, \{(\ref{EFE00})-(\ref{EFE22})\}, or \{(\ref{EFE00}),(\ref{EFE11}),(\ref{B-p-rho})\}; the EFE give three equations, but we have four unknown functions. Hence, another equation is needed to determine a definite solution. It is the approach to the choice of this extra equation that makes the works of Tolman \cite{Tolman} and Oppenheimer \& Volkoff \cite{OV} different.
   

\section{The Tolman approach}
\label{sect:Tolman}

  Tolman writes (where his $\lambda$ and $\nu$ are in lieu of our $A$ and $B$, respectively)
\begin{quotation}
  From a physical point of view, it might seem most natural to introduce this additional hypothesis in the form of an "equation of state" describing the relation between pressure p and density $\rho$ which could be expected to hold for the fluid under consideration. ...
  
  From a mathematical point of view, however, the derivatives of $\lambda$ and $\nu$ occur in our eqs. ... in such complicated and nonlinear manner that  we cannot in general expect to obtain explicit analytic solutions ... it proves more advantageous to introduce the additional equation necessary ... in the form of some relation, connecting $\lambda$ or $\nu$ or both with $r$, so chosen ... as to make the resulting set of equations readily soluble. 
 \end{quotation}
  
In particular, by setting $p$'s solved from (\ref{EFE11}) and (\ref{EFE22}) equal, one can get an equation involving $A$, $B$ and their derivatives (``equation of pressure isotropy''), and by making a choice for one of them, one might be able to solve for the other. $\rho(r)$ and $p(r)$ can then be trivially calculated by eqs.(\ref{EFE00}) and (\ref{EFE11}). Tolman calls this a ``mathematically rather than physically motivated procedure'' and remarks that it is not guaranteed that the solutions will be realistic (in the sense that the fluid properties that result are) and suggests that they might still be useful in understanding equilibrium conditions for actual fluids. He goes on to apply this technique to derive several solutions (Table \ref{table:Tolmans}). The (then) new solutions are, as expected, not very realistic.
\begin{table} [h!]
 \begin{tabular}{ | l | l  | p{70 mm} | } \hline
\multicolumn{1}{|c|}{{\bf Sol. name}} & \multicolumn{1}{c|}{{\bf Choice}} & \multicolumn{1}{c|}{{\bf Comment}}    \\  \hline \hline

Tolman I & $B$=const. & Einstein static universe (previously known) \\  \hline

Tolman II & $B/A$=const. &  Schwarzschild-de Sitter solution (previously known)  \\  \hline

Tolman III & $A = \frac{1}{1-r^{2}/R^{2}}$ & Schwarzschild interior solution (previously known)  \\  \hline

Tolman IV & $B'/2r$=const. & Finite sphere made of a fluid with a certain quadratic EoS, nonsingular $p(r)$ and $\rho(r)$   \\  \hline

Tolman V & $B$=const.$r^{2n}$ & Finite sphere made of a fluid with a certain nonsingular EoS, but with infinite central pressure and density \\  \hline 

Tolman VI & $A$=const. & Finite sphere made of a fluid with another certain nonsingular EoS, but with infinite central pressure and density \\  \hline

Tolman VII & $\frac{1}{A} = 1 - \frac{r^{2}}{R^{2}} + \frac{r^{4}}{D^{4}}$ & very complicated \\  \hline

Tolman VIII & $AB=$const.$r^{2b}$ & very complicated \\  \hline

\end{tabular}
\caption{Choices made by Tolman in his ``mathematical'' approach.}
\label{table:Tolmans}
\end{table}


\section{The OV approach}
\label{sect:OV}

In the Oppenheimer \& Volkoff paper, the approach called ``physical'' by  Tolman in the previous paper is adopted, that is, the additional equation is assumed to be an EoS for the fluid. As also stated by Tolman, this makes an analytical solution very difficult, if not impossible; but in the paper a procedure is set up for the attempt nevertheless, more importantly, the set-up comprises a very useful guide for {\it numerical} solutions for a given fluid, that is, a given EoS.

They are motivated by the integrability of the paranthesis in (\ref{EFE00}) to define a function $F(r)$
\begin{equation}
F(r) = \kappa \int \rho r^{2} dr          \label{FDef}
\end{equation}
which is $\kappa/4\pi$ times the total energy (mass) contained inside the radius $r$. Then one gets
\begin{equation}
A = \frac{r}{r-F},     \label{AinF}
\end{equation}
eq.(\ref{EFE11}) becomes
\begin{equation}
\frac{B'}{B} = \frac{\kappa p r^{2} + 1}{r-F}  -   \frac{1}{r}       \label{B'/B}
\end{equation}
and  finally substitution for $A$, $B$ and their derivatives in (\ref{EFE22}) or (\ref{B-p-rho}) gives
\begin{equation} 
p' = - \frac{(\kappa p r^{3} + F)}{2 r (r-F)} (\rho + p).     \label{OV}
\end{equation}

This is the equation that is usually called the OV or TOV equation. Sometimes eq.(\ref{FDef}) is included in the naming, then the plural is used.

To look for an analytical solution, in this equation one would write $p$ in terms of $\rho$ via an equation of state, then $\rho$ in terms of $F'$, via (\ref{FDef}), eventually getting a second order differential equation for $F$. After solving for $F$, $A$ and $B$ would be found via (\ref{AinF}) and (\ref{B'/B}), giving a metric for that equation of state. Unfortunately, the differential equations one gets for even the simplest EoS's (e.g. $p = w \rho$, with constant $w$) are usually impossible to solve analytically. One exception is $\rho$=const., which gives $F(r)$, and then from eq.(\ref{OV}) one gets a first order differential equation for $p(r)$, which eventually does give the solution for a sphere of constant density, i.e. the Schwarzschild interior solution.

But, the equations in this form are also very amenable to numeric treatment:
Assume that you know the values of $F(r)$ and $p(r)$ at some radius $r$. Then the density is given by the EoS, and the values of $F(r)$ and $p(r)$ at a slightly larger radius $r+\delta r$ can be found by eqs.(\ref{FDef}) and (\ref{OV}), respectively. Since $F(0)$ is necessarily zero for regular solutions, this means that one can start with a central density and work outward to numerically find $F(r)$ and $p(r)$, hence the line element. Note that this cannot be done with either of the sets \{(\ref{EFE00})-(\ref{EFE22}),EoS\} or \{(\ref{EFE00}),(\ref{EFE11}),(\ref{B-p-rho}),EoS\}.


\section{Conclusions}

To sum up, the Tolman \cite{Tolman} approach consists of facilitating an analytical solution by making a suitable  choice for (one of) the metric function(s), in the process usually sacrificing physical reasonability, because the choice has nothing to do with the physics, in fact, it dictates the physics.

The work of Oppenheimer \& Volkoff \cite{OV}, on the other hand, is an exploration of how to incorporate the physics of the perfect fluid, if known, into the solution. The equation (\ref{OV}) appears first in that paper, and not in Tolman's \cite{Tolman}; which is natural, since it is a product of the approach of that paper, which is in a sense diametrically opposite of that of Tolman\footnote{A few of the early users of the phrase ``TOV equation'', such as \cite{Wheeler68} and \cite{Calamai} cite Tolman's 1934 book \cite{TolmanBook}, but the equation does not appear there. In that book (Sect.95), the problem is set up, and the equivalent of eq.(\ref{B-p-rho}) is derived.}.

Hence, we conclude that the more appropriate name for the equation of hydrostatic equilibrium in a static spherically symetric spacetime sourced by a perfect fluid is ``The Oppenheimer-Volkoff (OV) equation''.

$\;$\\
$\;$\\

This research did not receive any specific grant from funding agencies in the public, commercial, or not-for-profit sectors.

\end{document}